\documentclass{PoS}
\title{Observations of the Crab Nebula with Early HAWC Data}

\ShortTitle{HAWC Crab Nebula Observations}

\author{\speaker{F. Salesa Greus} $^a$ for the HAWC Collaboration$^b$\\
        \llap{$^a$}Department of Physics, Pennsylvania State University, 16802 University Park, PA, USA\\
        \llap{$^b$}For a complete author list, see \href{http://www.hawc-observatory.org/collaboration/icrc2015.php}{www.hawc-observatory.org/collaboration/icrc2015.php}.\\
         E-mail: \email{sagreus@psu.edu}}
        
\abstract{The High Altitude Water Cherenkov (HAWC) Observatory is a TeV gamma-ray detector, completed in early 2015. HAWC started science operations in August 2013 with a third of the detector taking data. Several known gamma-ray sources have already been detected with the first HAWC data. Among these sources, the Crab Nebula, the brightest steady gamma-ray source at very high energies in our Galaxy, has been detected with high significance. In this contribution I will present the results of the observations of the Crab Nebula with HAWC, including time variability, and the detector performance based on early data.}

\FullConference{The 34th International Cosmic Ray Conference,\\
		30 July- 6 August, 2015\\
		The Hague, The Netherlands}

\begin{document}

\section{Introduction}%\input{intro}

The Crab Nebula is the brightest steady source in the known VHE gamma-ray sky. The first TeV gamma rays from the Crab Nebula were detected in the 80's by Whipple~\cite{whipple}. Since then, it has been extensively studied by gamma-ray experiments~\cite{HESS-Crab}~\cite{Milagro-Crab}.

The Crab Nebula has been detected at a >30$\sigma$ level in HAWC. %This significant detection can be used to compute the detector sensitivity. 
In this contribution we show the Crab Nebula observations for the two different datasets currently used in the HAWC analysis. We compare the measured excess with the expectations from simulation. We also show the Crab flux light curve to check for time variability. Finally with the current simulation we can derive our present detector sensitivity and compare it to the previous HAWC predictions and other experiments.

\section{Data selection}%\input{hardware}

HAWC is a gamma-ray observatory that consist of an array of water Cherenkov detectors (WCDs) spread on a 22,000~m$^{2}$ area on the slope of the Sierra Negra Volcano in Mexico\footnote{A more detailed description of the observatory can be found in~\cite{Andy}.}. HAWC construction ended in March 2015. However, the data collection started already two years before, producing the first scientific results~\cite{HAWC}. During the construction phase, HAWC was operated with a variable number of WCDs, ranging from $\sim$100 to almost 300 WCDs. In this contribution we will review the two main configurations. In the first data period, called HAWC111, the array grew from 106 to 133 live WCDs. This data set covers the period from August 2$^{nd}$ 2013 to July 7$^{th}$ 2014 for a total live-time of 283~days. The second period, HAWC250, where the number of operating WCDs ranged from 247 to 293, covers from November 26$^{th}$ 2014 to May 6$^{th}$ 2015. The total live-time for this period is 149~days.

Collected data were divided into 9 analysis bins (nHit bins) based on the number of triggered PMTs in the array in a particular event. The first nHit bin is defined for a given trigger rate of 5~kHz. The following bin edges are defined such that the rate decreases by a factor 2.

In order to reduce the overwhelming cosmic-ray background, we apply selection cuts to the reconstructed data to discriminate between gammas and hadrons~\cite{Andy}. A major effort is on-going for the improvement of the detector sensitivity by more efficient gamma/hadron separation~\cite{Tomas}~\cite{Zig}.

The selection cuts are different for each nHit bin, but are the same for a given HAWC configuration. The only exception is HAWC111 where we used two set of cuts due to hardware changes in the detector.
For the analysis shown here the selection cuts were optimized to maximize the data significance on the Crab. This choice intrinsically introduces a bias in the measured significance. In addition, quantities like the median energy or the angular resolution (AR) depend on the selection cuts, so they are different in each nHit bin. 
Therefore, the cuts used here are not necessarily optimum for other sources, e.g., extra-Galactic where a cut-off is expected at high energies. However, they can be considered a starting point for other sources.

In order to get the number of signal events we use circular angular bin centered at the position of the Crab Nebula.
The total number of events within the circular bin is compared to the expected background computed using the direct integration technique~\cite{Milagro-Crab}. We take the number of events in excess as our measured Crab signal. The significance was obtained with Eq. 17 in~\cite{LiMa}. The bin definition for each period and the value for each quantity are summarized in Tables~\ref{tb: table} and~\ref{tb: table2}. 

%------------- HAWC111 -------------------------------------------
\begin{table}[htpb]
\begin{center}
\small
%\footnotesize
%\tabcolsep=0.11cm
\begin{tabular}{|c|c|c|c|c|c|c|} \hline
{\bf Bin} & {\bf frac. NHit} & {\bf angular bin radius (deg)} & {\bf E (TeV)} & {\bf excess} & {\bf back} & {\bf signif}\\ \hline \hline
1 & 0.07-0.11 & 1.94 & 0.45 & 1.59~$\times$~10$^{4}$ & 1.14~$\times$~10$^{7}$ & 4.8\\\hline
2 & 0.11-0.16 & 1.48 & 0.65 & 4500 & 2.6~$\times$~10$^{6}$ & 4.4\\\hline
3 & 0.16-0.25 & 1.10 & 0.95 & 5000 & 5.1~$\times$~10$^{5}$ & 8.1\\\hline
4 & 0.25-0.37 & 0.95 & 1.5 & 1870 & 7.0~$\times$~10$^{4}$ & 9.0\\\hline
5 & 0.37-0.51 & 0.80 & 2.1 & 710 & 1.03~$\times$~10$^{4}$ & 10.0\\\hline
6 & 0.51-0.66 & 0.80 & 3.7 & 380 & 2400 & 9.1\\\hline
7 & 0.66-0.78 & 0.75 & 4.2 & 86 & 190 & 6.7\\\hline
8 & 0.78-0.88 & 0.45 & 5.9 & 13.9 & 7.1 & 4.8\\\hline
9 & 0.88-1.00 & 0.35 & 9.8 & 16.1 & 6.9 & 5.0\\\hline

\hline

\end{tabular}
\vspace{0.3 cm}
\caption{Summary of the 9 nHit bins for the first $\sim$180 days of HAWC111 analysis with information about fraction of PMTs hit, optimal circular bin radius, median energy, gamma-ray excess, expected background. Due to hardware changes the selection cuts were slightly modified for the last period of HAWC111. The last column shows the significance for the total 283 days included in the HAWC111 dataset.}
\label{tb: table}
\end{center}
\end{table}
%-------------------------------------------------------------

%--------------- HAWC250 ---------------------------------------
\begin{table}[htpb]
\begin{center}
\small
%\footnotesize
\begin{tabular}{|c|c|c|c|c|c|c|} \hline
{\bf Bin} & {\bf frac. NHit} & {\bf angular bin radius (deg)} & {\bf E (TeV)} & {\bf excess} & {\bf back} & {\bf signif}\\ \hline \hline
1 & 0.07-0.10 & 1.33 & 0.60 & 4900 & 9.6~$\times$~10$^{5}$ & 4.8\\\hline
2 & 0.10-0.16 & 0.93 & 0.94 & 3700 & 1.53~$\times$~10$^{5}$ & 9.2\\\hline
3 & 0.16-0.25 & 0.83 & 1.4 & 1900 & 2.5~$\times$~10$^{4}$ & 11.6\\\hline
4 & 0.25-0.36 & 0.7 & 2.3 & 1200 & 7200 & 13.5\\\hline
5 & 0.36-0.48 & 0.73 & 3.8 & 490 & 1550 & 11.6\\\hline
6 & 0.48-0.62 & 0.65 & 6.0 & 191 & 180 & 12.1\\\hline
7 & 0.62-0.74 & 0.55 & 9.8 & 79 & 32 & 10.6\\\hline
8 & 0.74-0.84 & 0.45 & 14 & 39 & 10.4 & 8.4\\\hline
9 & 0.84-1.00 & 0.4 & 24 & 13.5 & 1.53 & 6.2\\\hline

\hline

\end{tabular}
\vspace{0.3 cm}
\caption{Summary of the 9 nHit bins used in the HAWC250 with information about fraction of PMTs hit, optimal circular bin radius, median energy, gamma-ray excess, expected background, and significance.}
\label{tb: table2}
\end{center}
\end{table}
%----------------------------------------------------------

\begin{figure}[htpb]
\begin{center}
\includegraphics[width=0.7\linewidth, angle=0]{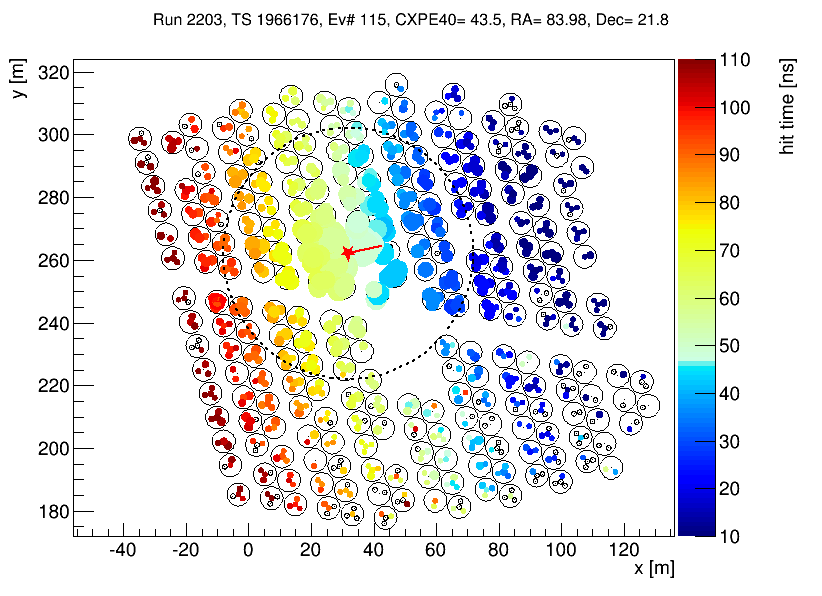}
\vspace{-0.1 cm}
\caption{An event detected with HAWC250. The reconstructed arrival direction is within 0.4$^{\circ}$ of the Crab Nebula. There were about 800 PMTs taking data at that time, 90\% of which were triggered in this event. The color scale corresponds to the trigger time from early in blue to late in red. The size of the circles is proportional to the cube root of the charge. The reconstructed shower core is marked with a red star.} 
\label{fig: Crab_event}
\end{center}
\end{figure}

The nHit bin number 9 is composed of events hitting most of the PMTs on the array. Therefore, it is the one with highest median energy, and where the gamma-hadron discrimination can be more efficient. In particular, for HAWC250 data, this bin is very gamma-rich with a ratio of 9:1 signal-background. Only a few events pass the cuts in this case because the flux for both cosmic-rays and Crab are much lower at high energy. Fig.~\ref{fig: Crab_event} shows one of those gamma-ray-like events with the HAWC event display. The colors are related to the arrival time of hits. The size of the circles is related to the deposited charge in the PMTs.

%%%%%%%%%%%%%%%%%%%%%%%%%%%
%%%%%%%%%%%%%%%%%%%%%%%%%%%

\section{Detection of the Crab Nebula}

For each period presented in the previous section, the 9 nHit bins were combined in a single significance map. Some assumptions were made to the combination. First, our simulated point spread function (PSF) was enlarged by an extra smearing in the arrival direction of the gamma rays to reproduce the measured AR. Then we assumed that our PSF is the result of the sum of two Gaussians.
Finally, we weight the bins according to their measured signal to background ratio as explained in~\cite{sensi-paper}. 

\begin{figure}[htpb]
\vspace{-0.7 cm}
\begin{center}
\begin{tabular}{c c}
\hspace{-0.8 cm}
\includegraphics[width=0.52\linewidth, angle=0]{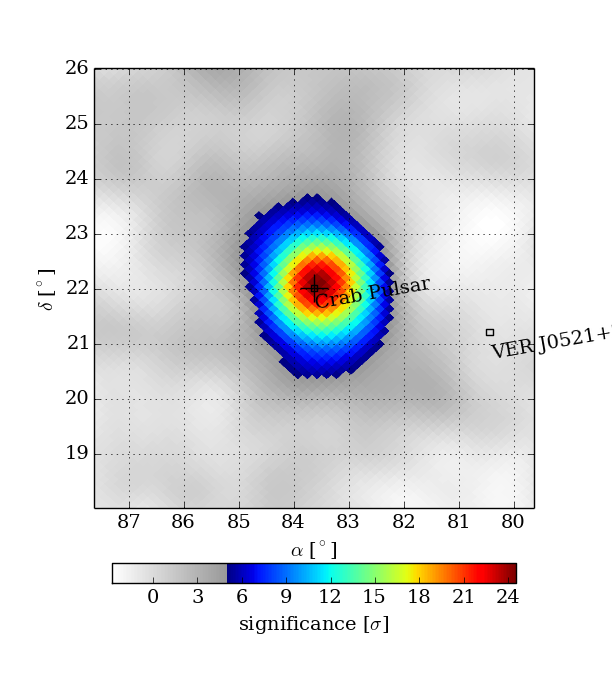}
\hspace{0.0 cm}
\includegraphics[width=0.52\linewidth, angle=0]{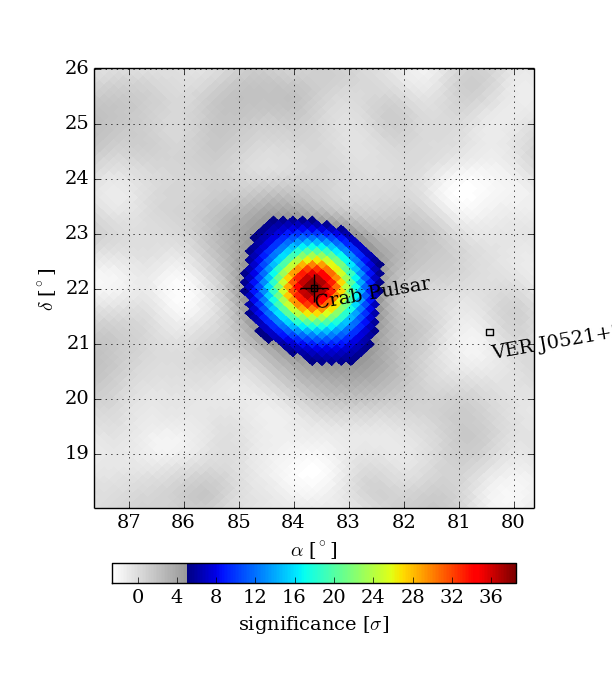}
\end{tabular}
\vspace{-0.5 cm}
\caption{Map of excess significances centered on the Crab Nebula in equatorial coordinates and Mercator projection using HAWC111 (left) and HAWC250 (right) data.}
\label{fig: crab}
\end{center}
\end{figure}

After these considerations, the observed significance at the Crab Nebula location is $\sim$24$\sigma$ for the HAWC111 time period (Fig.~\ref{fig: crab} left) and $\sim$38$\sigma$ for HAWC250 (Fig.~\ref{fig: crab} right). 

%%%%%%%%%%%%%%%%%%%%%%%%%%%
%%%%%%%%%%%%%%%%%%%%%%%%%%%
 
\section{Signal expectations}%\input{charge}

\begin{figure}[htpb]
\vspace{-1 cm}
\begin{center}
\begin{tabular}{c c}
\hspace{-0.2 cm}
\includegraphics[width=0.52\linewidth, angle=0]{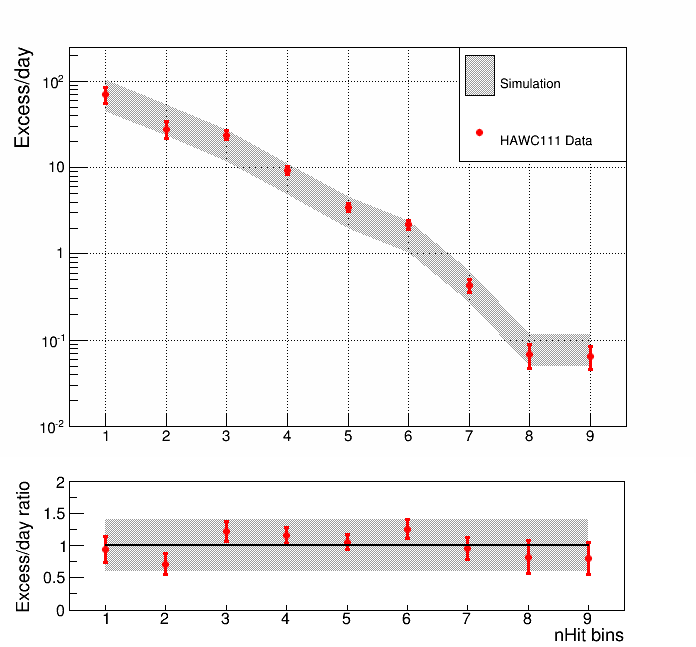}
\hspace{-0.2 cm}
\includegraphics[width=0.52\linewidth, angle=0]{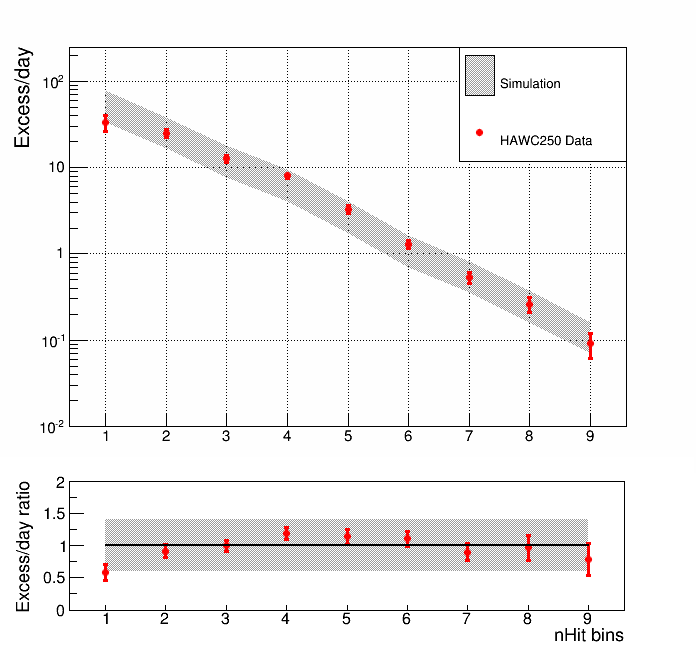}
\end{tabular}
\vspace{-0.1 cm}
\caption{Daily excess in data and simulations for the Crab nebula using the HAWC111 (left) and HAWC250 (right) data samples. The data error bars are statistical only. The median energies of each nHit bins are shown in Tables~\protect\ref{tb: table} and~\protect\ref{tb: table2}.}
\vspace{-0.5 cm}
\label{fig: excess}
\end{center}
\end{figure}

For each one of the nHit bins we compare the measured gamma-ray excess events with the values expected from simulation where a particular spectrum for the Crab gamma-rays is assumed (see Fig.\ref{fig: excess}). 
%%%%%%%%%%%%%%%%%%%%%%%%%%%
%%%%%%%%%%%%%%%%%%%%%%%%%%%

The circular angular bins used to count the number excess events in data are the ones listed in Tables~\ref{tb: table} and~\ref{tb: table2}. As we mentioned before, the measured AR is worse than predicted by simulations. We assume that the actual gamma rays arrive with wider PSF than predicted, but the excess rate is the same ($\sim$ 70\% in the optimal angular bin). To account for this we use a different circular angular bin in the simulations to compute the expected excess.

For the simulation we have used a 2-parameter power law function to model the flux:
%%%%%%%%%%%%%%%%%%%%%%%%%%%%%%%%%%                                                  
\begin{equation}
\frac{dN}{dE}(I_{0},\alpha) = I_{0} \bigg(\frac{E}{E_{0}}\bigg)^{-\alpha}
\label{eq: power-law}
\end{equation}
%%%%%%%%%%%%%%%%%%%%%%%%%%%%%%%%    
%\vspace{-2 cm}

\noindent where $I_{0}$ is the normalization flux, and $\alpha$ is the spectral index. The spectrum for the simulation was the one measured by H.E.S.S.~\cite{HESS-Crab}, i.e., $I_{0}$=3.45$\times10^{-11} cm^{-2}s^{-1}TeV^{-1}$ and $\alpha$=2.63.
The gray band reflects the 40\% systematic uncertainties explained in the section~\ref{sec: system}.

Data and the expectation from simulations agree within uncertainties.

\section{Systematic Uncertainties}
\label{sec: system}
The main sources of systematic uncertainties are: 

\begin{itemize}
   \item Uncertainties on the measured number of PEs based on how well we model the detector. In this regard, muon studies have shown that there is a scale discrepancy between the simulated charge and the charge from actual data. There are several causes that may explain this (e.g. the simulated quantum efficiency is higher than in data). Scaling the simulations to match the data and comparing the signal passing rates have shown that this effect is less than 20\%.
   
   \item Uncertainties coming from taking data at different stages of the detector, i.e., number of WCD/PMTs operational. This effect was studied for both periods by using different simulations assuming different number of PMTs. The discrepancy on the signal passing rates between them was below 20\%.
   
   \item AR uncertainties. We have used a circular angular bin to get the number of signal events. The size of the bin was chosen to maximize the significance on the Crab. Measuring our AR from data using a Gaussian PSF, we have seen that this choice may vary $\pm$20\% around the optimal angular bin.
   This uncertainty of $\pm$20\% in the bin size translates into 15-20\% uncertainty in the fraction of signal contained in the circular angular bin. 
   
  \end{itemize} 
 
 The three effects listed above gives an overall uncertainty on the expected excess of the order of 40\%.

\section{Time variability}%\input{timing}

Gamma-ray flares from the Crab Nebula have been reported by Fermi~\cite{Fermi-flare}. The procedure used in HAWC to look for flaring sources uses a likelihood method explained in~\cite{Robert}. The Crab flux light curve has been studied using this method. 

\begin{figure}[htpb]
\begin{center}
\includegraphics[width=0.8\linewidth, angle=0]{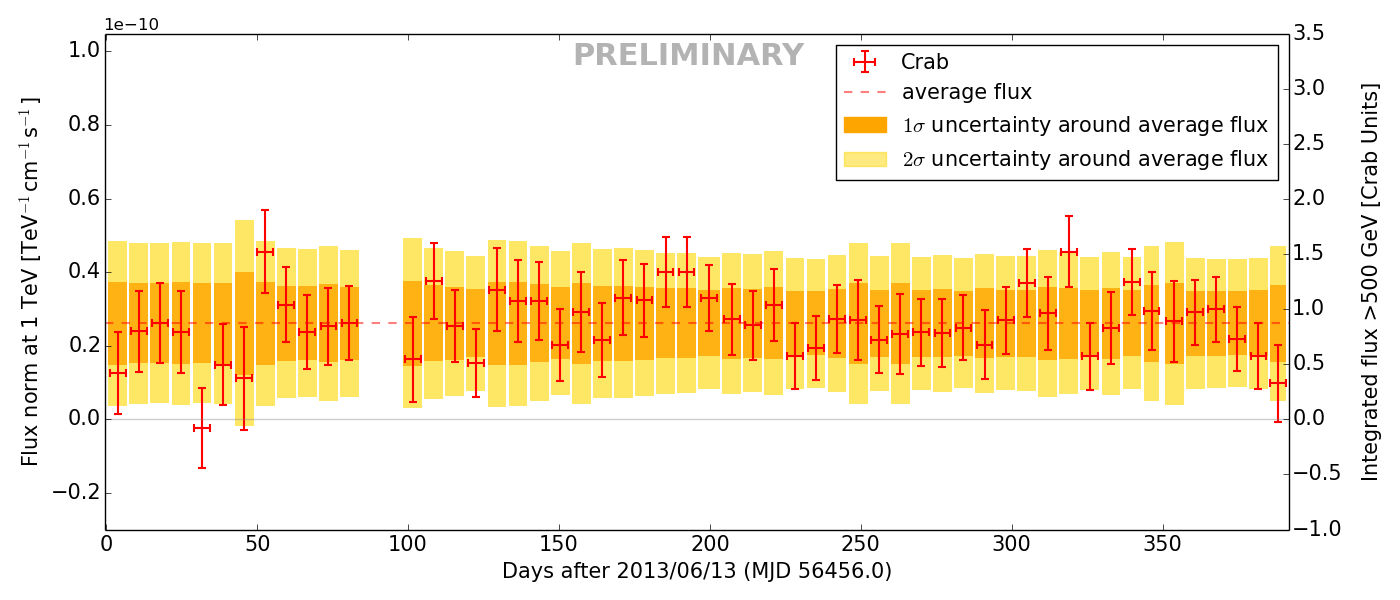}
\vspace{-0.1 cm}
\caption{HAWC flux light curve for the Crab, including data taken between June 13, 2013, and July 9, 2014, binned in 7-day intervals. The error bars are statistical only.}
\label{fig: Crab_time}
\end{center}
\end{figure}

Fig.~\ref{fig: Crab_time} shows the measured flux of the Crab Nebula in 7 days intervals, for the period between June 13, 2013, and July 9, 2014, HAWC111.
%%%%%%%%%%%%%%%%%%%%%%%%%%%
%%%%%%%%%%%%%%%%%%%%%%%%%%%
The left axis shows the normalization of the spectrum at 1 TeV. The right axis shows HAWC-based Crab units, defined by integrating the spectrum over energy above 500 GeV and dividing it by the integral of the Crab-reference simple power law spectrum with index 2.63 and normalization 2.99 cm$^{-2}$s$^{-1}$TeV$^{-1}$ (the average HAWC value for the 13-month time period). The gap in the plot is due to one week of detector maintenance performed during times when the Crab was transiting. The resulting low coverage of that week precludes a successful fit.

Following the prescription in~\cite{1FGL} to test the variability on the Crab light curve, we got a p-value of 0.91.
Therefore, no significant flares from the Crab Nebula were detected with HAWC111. The HAWC250 data is presently being analyzed.

\section{Detector sensitivity}%\input{performance}

\begin{figure}[htpb]
\begin{center}
\includegraphics[width=0.75\linewidth, angle=0]{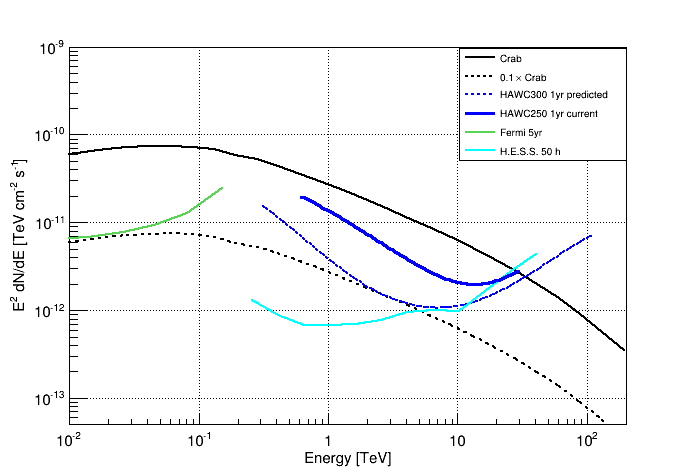}
\vspace{-0.1 cm}
\caption{Differential sensitivity of a $5\sigma$ discovery in one year with HAWC. The predicted sensitivity of HAWC300 comes from the studies performed in~\cite{sensi-paper}. The sensitivity of other experiments is also shown for comparison.}
\label{fig: sensi}
\end{center}
\end{figure}

It has been shown that the excesses detected from the Crab in each of the nHit bins are within uncertainties. We can use our simulation to estimate the required flux from a source for a 5$\sigma$ discovery.
We simulated a source with a Crab-like spectrum transiting at a declination of 35 degrees, which is the typical declination for a source in the HAWC (latitude $19.0^{\circ}$N) field of view.
Fig.~\ref{fig: sensi} shows the HAWC 5$\sigma$ differential sensitivity per quarter decade in energy, i.e., bins of 0.25 in log10 space. The sensitivity is presented in the plot as the fit to a third degree polynomial of the nHit bins used in our simulation.
%%%%%%%%%%%%%%%%%%%%%%%%%%%
%%%%%%%%%%%%%%%%%%%%%%%%%%%

\section{Conclusions}
The Crab Nebula has been detected with high significance both in the HAWC111 and HAWC250 datasets. The observed signal from the Crab is reproduced by the simulations. 

From the flaring analysis there is no evidence for the Crab Nebula emitting significantly higher with respect to its quiescent flux during the HAWC111 period.

We have used the Crab Nebula as a test source to compute the 5$\sigma$ differential sensitivity of the detector at the present stage. A comparison with the previous predictions and other gamma-ray experiments has been presented.
The current sensitivity for HAWC250 is about factor 2 lower than what is expected for HAWC250, and almost a factor 3 compared to what was reported for HAWC300 in~\cite{sensi-paper}. The analysis is still in a preliminary stage. We still need a better understanding of our instrument response. Our predicted PSF did not reproduce well what is observed in data. The PSF correction we did can by itself account for more than half of the discrepancy between the predicted sensitivity and the current sensitivity for HAWC250. Moreover, we are working on improving the detector calibration as well as the gamma/hadron selection. All these effects combined are expected to improve the detector sensitivity to at least the previously reported expectations.

\section*{Acknowledgments}
\footnotesize{
We acknowledge the support from: the US National Science Foundation (NSF);
the US Department of Energy Office of High-Energy Physics;
the Laboratory Directed Research and Development (LDRD) program of
Los Alamos National Laboratory; Consejo Nacional de Ciencia y Tecnolog\'{\i}a (CONACyT),
Mexico (grants 260378, 55155, 105666, 122331, 132197, 167281, 167733);
Red de F\'{\i}sica de Altas Energ\'{\i}as, Mexico;
DGAPA-UNAM (grants IG100414-3, IN108713,  IN121309, IN115409, IN111315);
VIEP-BUAP (grant 161-EXC-2011);
the University of Wisconsin Alumni Research Foundation;
the Institute of Geophysics, Planetary Physics, and Signatures at Los Alamos National Laboratory;
the Luc Binette Foundation UNAM Postdoctoral Fellowship program.
}

\end{document}